\documentclass[conference]{IEEEtran}
\IEEEoverridecommandlockouts

\usepackage{graphicx}
\usepackage{multirow}
\usepackage{caption}
\usepackage{subcaption}
\usepackage{lipsum} 
\usepackage{listings}
\usepackage{balance}
\usepackage{cite}
\usepackage{amsmath,amssymb,amsfonts}
\usepackage{algorithmic}
\usepackage{textcomp}
\usepackage{xcolor}
\usepackage{colortbl}
\usepackage{float}
\usepackage{enumitem}
\usepackage{framed}
\usepackage{comment}
\def\BibTeX{{\rm B\kern-.05em{\sc i\kern-.025em b}\kern-.08em
    T\kern-.1667em\lower.7ex\hbox{E}\kern-.125emX}}
    
\definecolor{codepurple}{rgb}{0.58,0,0.82}
\definecolor{backcolour}{RGB}{239, 239, 239}
\definecolor{codeorange}{RGB}{191, 94, 45}
\definecolor{codeblue}{RGB}{0, 0, 255}
\definecolor{codegreenl}{rgb}{0,0.6,0}
\definecolor{codegreend}{RGB}{101, 139, 111}
\definecolor{codegray}{rgb}{0.5,0.5,0.5}

\lstdefinestyle{mystyle}{
    backgroundcolor=\color{backcolour},   
    commentstyle=\color{codepurple},
    keywordstyle=\color{codeorange},
    numberstyle=\small\color{codegray},
    stringstyle=\color{codegreenl},
    basicstyle=\fontsize{8}{8}\selectfont\ttfamily,
    breakatwhitespace=false,     
    breaklines=true,                 
    captionpos=b,                    
    keepspaces=true,                 
    numbers=none,                    
    numbersep=2pt,        
    numbers=left,
    stepnumber=1,
    showspaces=false,                
    showstringspaces=false,
    showtabs=false,                  
    tabsize=2
}
\begin{document}

\title{ProvLight: Efficient Workflow Provenance Capture on the Edge-to-Cloud Continuum}

\author{\IEEEauthorblockN{
Daniel Rosendo\IEEEauthorrefmark{1},
Marta Mattoso\IEEEauthorrefmark{2},
Alexandru Costan\IEEEauthorrefmark{1},
Renan Souza\IEEEauthorrefmark{3},
Débora Pina\IEEEauthorrefmark{2},\\
Patrick Valduriez\IEEEauthorrefmark{4},
Gabriel Antoniu\IEEEauthorrefmark{1}}
\IEEEauthorblockA{\IEEEauthorrefmark{1}University of Rennes, Inria, CNRS, IRISA - Rennes, France\\
\{daniel.rosendo, alexandru.costan, gabriel.antoniu\}@inria.fr}
\IEEEauthorblockA{\IEEEauthorrefmark{2}Federal University of Rio de Janeiro, Brazil\\
\{marta, dbpina\}@cos.ufrj.br}
\IEEEauthorblockA{\IEEEauthorrefmark{3}Oak Ridge National Laboratory, USA\\
souzar@ornl.gov}
\IEEEauthorblockA{\IEEEauthorrefmark{4}University of Montpellier, Inria, CNRS, LIRMM - Montpellier, France\\
patrick.valduriez@inria.fr}
}


\maketitle


\begin{abstract}
Modern scientific workflows require hybrid infrastructures combining numerous decentralized resources on the IoT/Edge interconnected to Cloud/HPC systems (aka the \emph{Computing Continuum}) to enable their optimized execution. Understanding and optimizing the performance of such complex Edge-to-Cloud workflows is challenging. 
Capturing the provenance of key performance indicators, with their related data and processes, may assist in understanding and optimizing workflow executions. 
However, the capture overhead can be prohibitive, particularly in resource-constrained devices, such as the ones on the IoT/Edge.


To address this challenge, based on a performance analysis of existing systems, we propose ProvLight, a tool to enable efficient provenance capture on the IoT/Edge. We leverage simplified data models, data compression and grouping, and lightweight transmission protocols to reduce overheads. We further integrate ProvLight into the E2Clab framework to enable workflow provenance capture across the Edge-to-Cloud Continuum. This integration makes E2Clab a promising platform for the performance optimization of applications through reproducible experiments.

We validate ProvLight at a large scale with synthetic workloads on 64 real-life IoT/Edge devices in the FIT IoT LAB testbed. Evaluations show that ProvLight outperforms state-of-the-art systems like ProvLake and DfAnalyzer  
in resource-constrained devices.
ProvLight is 26---37x faster to capture and transmit provenance data; uses 5---7x less CPU; 2x less memory; transmits 2x less data; and consumes 2---2.5x less energy. ProvLight~\cite{git-provlight} and E2Clab~\cite{e2clab-code} are available as open-source tools.
\end{abstract}


\section{Introduction}
\label{sec:introduction}

%






Data processing and Artificial Intelligence (AI) workflows can no longer rely on traditional approaches (due to resource usage, latency, and privacy constraints)~\cite{rosendo:hal-03654722} that send all data to centralized and distant Cloud datacenters for processing or AI model training~\cite{deng2020edge}. Instead, they need to leverage hybrid (decentralized) approaches that take advantage of the numerous resources close to the data generation sites (\emph{i.e.}, on the edge of the network)  to promptly extract insights~\cite{asch2018big} and satisfy the ultra-low latency requirements of applications. 

This hybrid approach contributes to the emergence of the \emph{Computing Continuum}~\cite{etp4-hpc-20} (or the \emph{Edge-to-Cloud Continuum} or the \emph{Transcontinuum}). It seamlessly combines resources and services at the center of the network (\emph{e.g.}, in Cloud datacenters), at its edge, and \emph{in-transit}, along the data path. Typically, data is first generated and preprocessed (\emph{e.g.}, model training with local data) on IoT/Edge devices. Then, data is transferred to (HPC-enabled) Clouds for Big Data analytics, AI model training, and global simulations. For instance, in Federated Learning (FL) model training, a central Cloud server collects data (model updates) from multiple decentralized Edge devices, then it generates a single accurate global inference model.

Due to the complexity incurred by application deployments on such highly distributed and heterogeneous Edge-to-Cloud infrastructures, realizing the Computing Continuum vision in practice is challenging. Deploying, analyzing, and reproducing performance trade-offs and optimizing large-scale, real-life applications on such infrastructures is difficult~\cite{rosendo:hal-03654722}. It requires configuring a myriad of system-specific parameters (\emph{e.g.,} from AI and Big Data systems) and reconciling many requirements or constraints in terms of energy consumption, network efficiency, and hardware resource usage, to cite a few~\cite{xia2018combining}. In recent works, these challenges have been mainly explored by systems like Pegasus~\cite{tanaka2022automating}, E2Clab~\cite{rosendo:hal-03310540}, Delta~\cite{kumar2021coding}.

The process of understanding, optimizing, and reproducing complex Edge-to-Cloud workflows may be assisted by \textbf{provenance data capture}. "Provenance data" refer to a record trail that accounts for \textbf{the origin of a piece of data} together with descriptions of the computational processes that assist in explaining \textbf{how and why it was generated}~\cite{liu2009encyclopedia}. \textbf{Capturing provenance data during workflow execution} helps users in tracking inputs, outputs, and processing history, allowing them to steer workflows precisely~\cite{souza_keeping_2019}.

For instance, considering a Federated Learning model training workflow executed on distributed devices on the Edge, the captured data during model training helps answer questions like: \textit{(i) What are the elapsed time and the training loss in the latest epoch for each hyperparameter combination?~\cite{silva2021capturing, asouza2020workflow}} or \textit{(ii) Retrieve the hyperparameters which obtained the 3 best accuracy values for model \emph{m}?~\cite{pina2021provenance,asouza2020workflow}}. Answering such queries helps to analyze hyperparameter values related to the training stages and to adjust them for better-quality results.

\subsection{Challenges and Novelty}

Overhead in provenance systems is a critical problem that must be assessed~\cite{herschel2017survey}. Many other contributions in provenance systems evaluate the overhead, such as~\cite{souza2019efficient, silva2020dfanalyzer}. Overhead is even more critical in edge devices because of resource constraints and power consumption. For this reason, we decided to focus on evaluating overhead in our work. In~\cite{balazinska2020next}, leading database researchers discussed the challenges of deploying services considering disaggregation and high heterogeneity of resources in hybrid cloud infrastructures. In~\cite{da2023workflows}, the authors describe challenges related to capturing provenance on the Edge-to-Cloud Continuum.

The main state-of-the-art provenance systems were designed to run on Cloud/HPC infrastructures. We highlight that we have not found in the literature reference systems tailored for IoT/Edge devices. Therefore, this work refers to systems well-known for their low provenance capture overhead in Cloud/HPC, such as DfAnalyzer~\cite{silva2020dfanalyzer}, ProvLake~\cite{souza2019efficient}, and PROV-IO~\cite{han2022prov}. We also include Komadu~\cite{tas2016approach} in our analysis, as it is also compared within the aforementioned works.

Enabling provenance data capture with low overhead in resource-constrained IoT/Edge devices \textbf{cannot be easily achieved} by existing provenance systems, calling for practical solutions beyond the state-of-the-art. For instance, it requires the design and development of novel capture approaches focusing on the hardware limitations of IoT/Edge devices, as proposed in this work.

\subsection{Contributions}

We make the following contributions: 

\begin{enumerate}

    \item The first research question we aim to answer is: \textbf{\emph{How Do the Existing Provenance Systems Perform in IoT/Edge Devices?}} We address this research question by providing an experimental evaluation of existing provenance systems along with a detailed discussion in Section~\ref{sec:soalimitations}.

    \item As our experiments concluded that the state-of-the-art systems present high overheads to capture provenance data in IoT/Edge devices, \textbf{we propose a novel workflow provenance data capture approach tailored for resource-limited IoT/Edge devices}, that addresses the limitations 
    found in the state of the art (Section~\ref{sec:provlight}). \textbf{ProvLight is an open-source implementation} of this approach (available at~\cite{git-provlight}), following the W3C PROV-DM recommendations. 
    
    \item \textbf{An integration of ProvLight within the E2Clab} automatic deployment and performance optimization framework. This \textbf{enables provenance data capture across the Computing Continuum} for hybrid workflows deployed on both IoT/Edge  and Cloud/HPC infrastructures. To the best of our knowledge, this enhanced version of E2Clab is \textbf{the first framework to support the end-to-end provenance data capture} of complex workflows executed on the Edge-to-Cloud Continuum (Section~\ref{sec:provlight_e2clab_extension}). This integration with E2Clab is an open-source tool available at~\cite{e2clab-code}. We highlight that, \textbf{ProvLight may be easily integrated into other deployment and performance optimization systems/frameworks}. 

    \item \textbf{A large-scale experimental validation} of ProvLight with synthetic workloads \textbf{on 64 real-life IoT devices} (from the FIT IoT LAB~\cite{adjih2015fit} testbed) and \textbf{Cloud resources} (from the Grid'5000~\cite{RaphaEtAl2006} testbed). Experimental evaluations show that ProvLight outperforms (\emph{i.e.,} lower capture overhead) DfAnalyzer and ProvLake systems in terms of capture time, CPU and memory usage, network usage, and power consumption (Section~\ref{sec:evaluation}).
    
\end{enumerate}

\section{Background}
\label{sec:background}



This work focuses on provenance systems \textbf{leveraging the user-defined capture approach}~\cite{khoi2022proml, spinuso2019active}. This approach allows users to define what to capture by workflow script instrumentation through capture libraries. 
We highlight that script instrumentation (\textit{e.g.}, adding logging calls) is a common practice in distributed systems, particularly to assist debugging. In provenance capture, many other approaches rely on script instrumentation \cite{xu2020survey, souza2019efficient, spinuso2019active}.
A good practice to promote data interoperability is that such libraries should follow provenance specifications like the PROV-DM recommendation, as an example. Finally, the provenance capture of Edge-to-Cloud workflows is a new topic that requires automatic deployment tools like E2Clab. 

\begin{figure}[t]
  \centering
  \includegraphics[width=\linewidth]{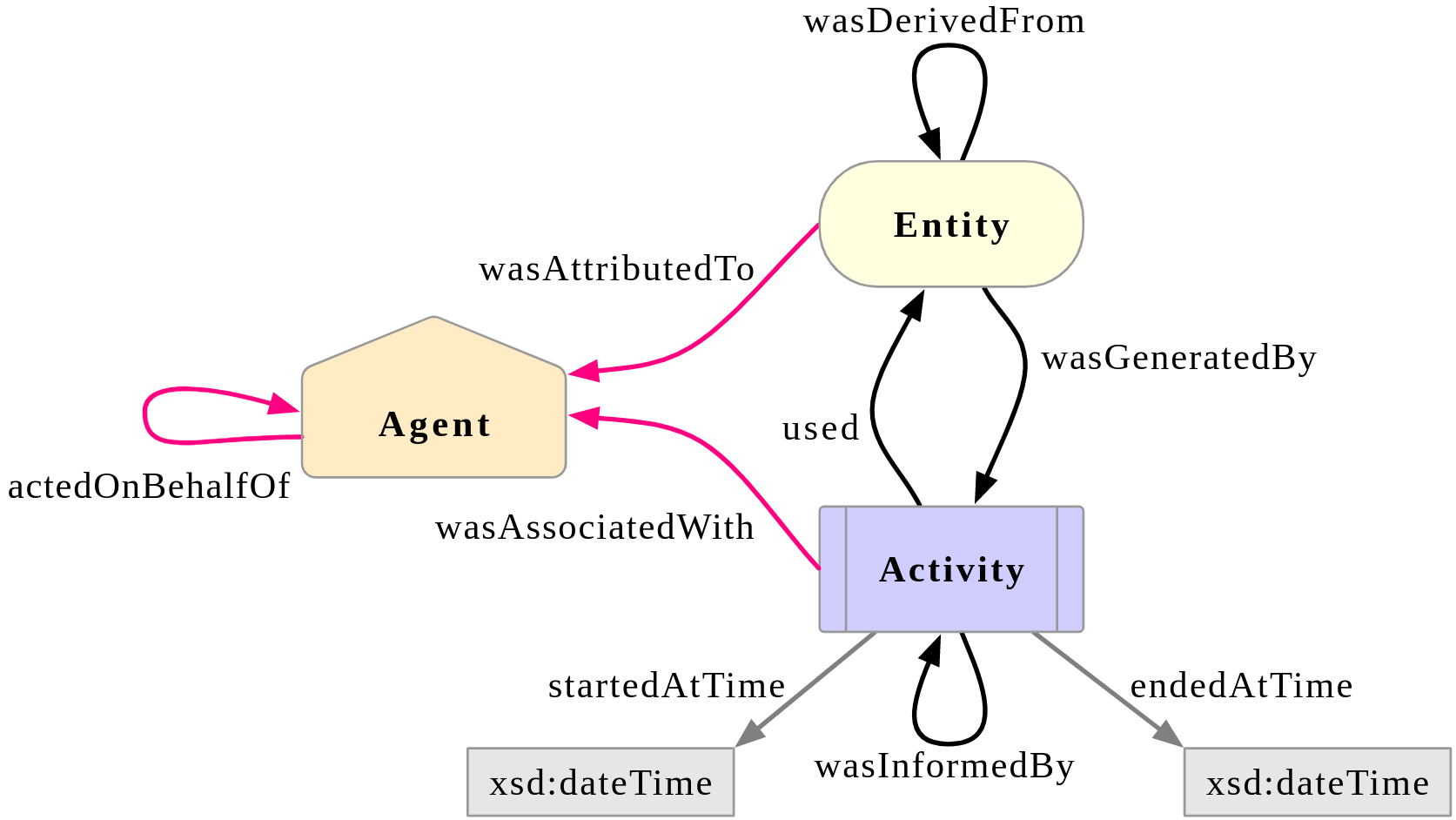}
  \caption{PROV-DM: The W3C PROV Data Model~\cite{belhajjame2013prov}.}
  \label{fig:w3c_provdm}
\end{figure}

\subsection{PROV-DM: The PROV Data Model}

PROV~\cite{missier2013w3c} is a specification to interchange provenance information. PROV-DM~\cite{belhajjame2013prov} is the data model for the W3C provenance family of specifications. It aims to promote data interoperability from provenance management systems. Provenance systems such as DfAnalyzer~\cite{silva2020dfanalyzer}, ProvLake~\cite{souza2019efficient}, PROV-IO~\cite{han2022prov}, Komadu~\cite{tas2016approach}, among many others, follow the PROV-DM model. 

Figure~\ref{fig:w3c_provdm} illustrates the core elements of PROV-DM and their relationships. PROV-DM provides an abstract representation of provenance data derivations. Briefly described, the core elements are: \emph{(i) Agent:} refers to tools invoked on behalf of users (\emph{e.g.,} software); \emph{(ii) Activity:} refers to tasks (\emph{e.g.,} processing steps); and \emph{(i) Entity:} refers to data objects (\emph{e.g.,} files, input parameters, etc.). Our capture approach also follows the PROV-DM recommendation.

\subsection{Capturing Provenance for Edge-to-Cloud Workflows}

\subsubsection{\textbf{Edge-to-Cloud Computing Continuum}} Edge infrastructures refer to computing and storage resources located where the data originated. They consist of numerous smart devices sensing \textit{"what"} is happening in the environment and generating potentially huge data streams at potentially high rates~\cite{rosendo:hal-03654722}. The Edge computing paradigm aims to push intelligence to those devices and extract value from data in real-time to improve response times while preserving privacy and security (critical data is analyzed locally).

Cloud infrastructures provide virtually "unlimited" computing and storage resources used essentially for backup and data analytics for global insight extraction in centralized data centers. Data is first ingested at high rates through dedicated systems (\emph{e.g.,} Apache Kafka\cite{garg2013apache}) and analyzed by Big Data processing frameworks (\emph{e.g.,} Spark\cite{apache-spark}). They perform stream and batch analytics on vast historical data, AI model training, and complex simulations. The goal is to help understand \textit{"why"} the phenomena sensed at the Edge are happening. 

\subsubsection{\textbf{Federated Learning Training Use Case}} To illustrate an Edge-to-Cloud application workflow, we refer to Federated Learning model training. Federated Learning~\cite{mcmahan2017communication} is a collaborative machine learning paradigm that trains a centralized model on decentralized and private data.


The Federated Learning architecture is composed of a central server (typically deployed on the Cloud) and various devices (deployed on the Edge). Edge devices first download a global model from the cloud server and train it for several epochs with their local data. After multiple rounds of model updates, the results are sent to the cloud server for global model aggregation. This training loop continues until the global model achieves the desired accuracy~\cite{ye2020edgefed}.

Capturing provenance data of Federated Learning model training at runtime helps scientists to track model training inputs (\emph{e.g.,} hyperparameters), outputs (\emph{e.g.,} accuracy), and processing history (\emph{e.g.,} training epochs). In this context, captured data from each training epoch may refer to the hyperparameters (input data) followed by the respective accuracy obtained from the training (output data). The goal is to allow users to answer queries like the ones presented in  Section~\ref{sec:introduction}). Analyzing hyperparameters along the model training allows for adapting the training data and fine-tuning the model. Provenance data traces also help in the interpretation and reproducibility of the training results~\cite{pina2020provenance,asouza2020workflow}.

\subsection{E2Clab: Reproducible Edge-to-Cloud Experiments}

\begin{figure}[t]
    \centering
         \includegraphics[width=\linewidth]{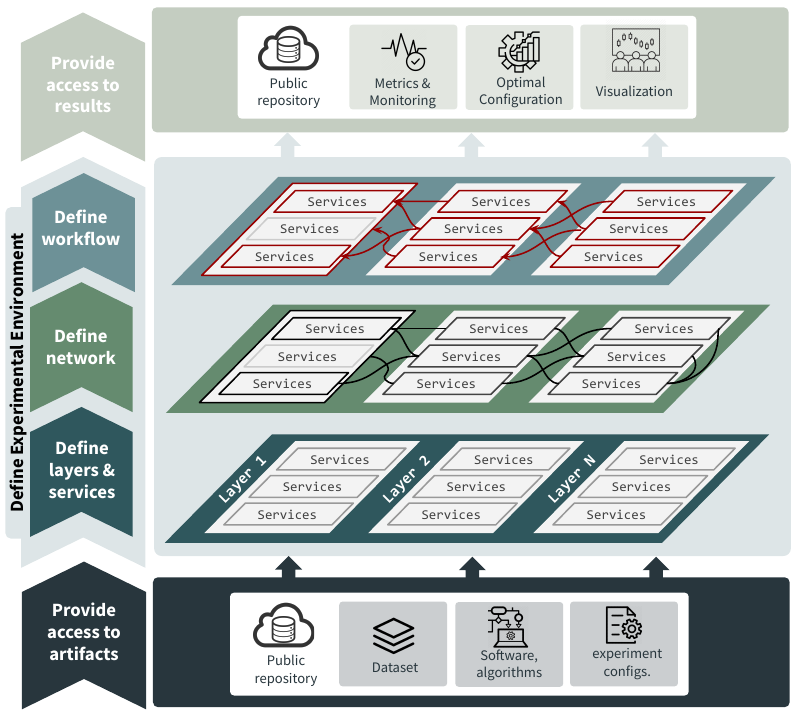}
    \caption{E2Clab experiment methodology~\cite{rosendo:hal-02916032}.}
    \label{fig:methodology}
\end{figure}

E2Clab~\cite{rosendo:hal-02916032} is an open-source framework (available at~\cite{e2clab-code}) that allows researchers to reproduce the application behavior in a controlled environment to understand and to optimize performance~\cite{rosendo:hal-03310540}. It sits on top of EnOSlib~\cite{cherrueau:hal-03324177} and implements a rigorous methodology (illustrated in Figure~\ref{fig:methodology}) for designing experiments with real-world workloads on the Edge-to-Cloud Computing Continuum. Section~\ref{sec:provlight_e2clab_extension} details how we extend E2Clab to enable provenance data capture of Edge-to-Cloud workflows. Figure~\ref{fig:e2clab-prov} illustrates the extended architecture.

High-level features provided by E2Clab are \emph{(i)} reproducible experiments; \emph{(ii)} mapping application parts (executed on Edge, Fog, and Cloud/HPC) and physical testbeds; \emph{(iii)} experiment variation and transparent scaling of scenarios; \emph{(iv)} defining Edge-to-Cloud network constraints; \emph{(v)} automatic experiment deployment, execution, and monitoring (\emph{e.g.,} on various testbeds like Grid'5000~\cite{RaphaEtAl2006}, Chameleon~\cite{KateEtAl2020}, and FIT IoT LAB~\cite{adjih2015fit}); \emph{(vi)} workflow optimization.

\section{Provenance Systems in IoT/Edge}
\label{sec:soalimitations}

In the literature, to the best of our knowledge, we have not found provenance data capture tools tailored for IoT/Edge devices. Capturing provenance data in such devices requires using tools designed for Cloud/HPC resources. Therefore, in this section, we assess their overhead for capturing data in resource-limited computing resources.

\subsection{Experimental Setup}
\label{subsec:setup}

\paragraph{\textbf{Selected provenance systems.}} 
Due to the limitations of the PROV-IO and Komadu systems, shown in Table~\ref{tbl:limitations}, they were excluded from our performance analysis. We choose ProvLake and DfAnalyzer because we have access to their data capture components as open-source software. Since we are limited to testing with the open-source version of these systems, we cannot experiment with features that might deliver lower overhead but are not open-source. For instance, ProvLake reports being able to use a different communication protocol other than HTTP 1.1 for machine learning provenance capture with low overhead in an HPC environment \cite{asouza2020workflow}, but this system version is not available as open-source.

\paragraph{\textbf{Performance metrics.}} 
The main analyzed metric is the \emph{capture time overhead}, which refers to the relative difference of the workflow execution time with and without data capture. 
We repeat the experiment 10 times for each provenance system and for each synthetic workload and report the mean followed by the 95\% confidence interval.

\paragraph{\textbf{Overhead levels.}} 
In the literature, the reference to \emph{low overhead} or \emph{negligible overhead}, in terms of provenance capture time in Cloud/HPC environments, differs between application domains. For instance: $<$2\% for blockchains~\cite{ruan2021lineagechain}; $\leqslant{}$4\% for I/O-centric workflows~\cite{han2022prov}; 4\% for AI model training~\cite{silva2021capturing}; $\leqslant{}$12\% for security applications~\cite{pasquier2018runtime}; to cite a few. Regarding provenance capture on resource-limited IoT/Edge devices, prohibitive overhead levels may vary depending on the application use case. For instance, in latency-sensitive applications such as autonomous vehicles~\cite{de2022trade}, real-time monitoring in smart energy grids~\cite{abir2021iot}, and virtual and augmented reality~\cite{lo2018edge}, to cite a few, a $>3\%$ processing time overhead is considered high (\emph{i.e.,} enough to exceed the acceptable latency thresholds) as it can introduce delays that disrupt the real-time nature of the application, leading to inaccuracies, missed targets, or compromised safety.


\begin{table}[t]
\centering
\small
\caption{Synthetic workload configurations.}
\label{tbl:workload_conf}
\begin{tabular}{rllll}
\hline
\multicolumn{5}{c}{
\begin{tabular}[c]{@{}c@{}}
\textbf{Configurations to generate the} \textbf{synthetic workloads}
\end{tabular}
}                                                                                        \\ \hline
\cellcolor[HTML]{CFE2F3}\begin{tabular}[c]{@{}r@{}}Number of chained transformations\end{tabular} & \cellcolor[HTML]{CFE2F3}5   &                             &     &   \\ \cline{1-2}
\cellcolor[HTML]{FCE5CD}Number of tasks                                                                   & \cellcolor[HTML]{FCE5CD}100 &                             &     &   \\ \cline{1-3}
\cellcolor[HTML]{FFF2CC}Attributes per task                                                              & \cellcolor[HTML]{FFF2CC}10  & \cellcolor[HTML]{FFF2CC}100 &     &   \\ \hline
\rowcolor[HTML]{D9D2E9} 
Task duration (s)                                                                                         & 0.5                         & 1                           & 3.5 & 5 \\ \hline
\end{tabular}
\end{table}

\paragraph{\textbf{Synthetic workload.}}

We use a synthetic workload to evaluate the provenance capture overhead because doing it in real workloads is much more complicated, costly, and may not make sense for the real application. The reason is that we cannot precisely control and isolate variables such as elapsed time, number of tasks, and number of attributes. A similar situation happens when scientists need to rely on simulations instead of real phenomena to test and evaluate their hypotheses. Unfortunately, there are no well-established benchmarks in the community to evaluate overhead in provenance systems. Therefore, like related work~\cite{souza2019efficient, silva2020dfanalyzer}, we decided to focus our analysis on synthetic workload configurations. Such configurations are based on real-life workloads~\cite{zhang2021federated, liu2020deep, ito2021device}, and we refined the configuration space of our workloads with preliminary experiments on real-life edge devices.

Table~\ref{tbl:workload_conf} presents the 8 synthetic workload configurations used to analyze the data capture overhead. We chose these values to cover combinations of application characteristics. The idea of these configurations is to mimic the characteristics of the various real-life workloads that IoT/Edge devices typically execute, such as AI model training (\emph{e.g.,} the Federated Learning use case we presented earlier), image pre-processing, and sensor data aggregation, among others. Such workloads are composed of various tasks (number of tasks), each one with a different number of attributes (attributes per task) and with different processing times (task duration).\\



We consider workloads with 5 chained transformations, which is an approximate number of transformations in many applications. In the Federated Learning application, for example, one of the transformations is model training, which has many epoch executions. We consider each epoch execution as a task of the model training transformation and each epoch has associated features (considered input attributes) and performance metrics (considered output attributes)~\cite{asouza2020workflow}. Other transformations include data preparation and the evaluation of the trained model.
To generate our synthetic workload, we consider 100 tasks.
In the Federated Learning example, it would represent a training with 100 epochs.
For each task, we represent applications that manipulate a few (about 10) or more (about 100) attributes per task.
Besides, to represent various classes of applications, we also consider four different task duration: shorter (\emph{e.g.,} 0.5 or 1 seconds) and longer (\emph{e.g.,} 3.5 or 5 seconds). 

We run \textbf{preliminary experiments to refine} the synthetic workload configurations. We observe that there is no significant impact on the capture overhead when varying the \emph{number of tasks} from 10, 50 to 100. In addition, since the data capture and transmission is measured per task, \textbf{mainly variations in the number of} \emph{attributes per task} (amount of data transmitted) and \emph{task duration} (data capture frequency) impact the capture time overhead (calculated as the relative difference).


\paragraph{\textbf{Hardware.}} 
Each workload configuration runs on a single A8-M3~\cite{a8m3iotlab} IoT device (ARM Cortex-A8 microprocessor, 600Mhz, 256MB; radio: 802.15.4, 2.4 GHz; power: 3.7V LiPo battery, 650 mAh) available at the FIT IoT LAB testbed~\cite{adjih2015fit}. We instrument the synthetic workloads (code available at~\cite{exp-artifacts}) with the capture libraries provided by ProvLake and DfAnalyzer systems. The libraries transmit the  data to the provenance system running on a remote Cloud/HPC server~\cite{grosg5k} (Intel Xeon Gold 5220, 2.20GHz, 18 cores; 96GB RAM; Ethernet) available at the Grid'5000~\cite{RaphaEtAl2006} testbed.

\subsection{Overhead Analysis} 
Table~\ref{tbl:overhead_existing_solutions} presents the \textbf{capture time overhead} of ProvLake and DfAnalyzer \textbf{in IoT/Edge devices}, and Table~\ref{tbl:provlake_grouping} shows the analysis of a feature provided by ProvLake, which consists of \textbf{grouping the captured data}, \emph{i.e.,} messages, before transmitting them to the server, \emph{i.e.,} provenance system. In addition, we analyze how low-bandwidth networks may impact such data grouping strategy. 

Results in Table~\ref{tbl:overhead_existing_solutions} show that both systems present high overhead ($>$39\%) for tasks with a duration of 0.5 seconds. For the remaining task duration, the overhead is still high ($>$3\%). Varying the number of attributes per task from 10 to 100 slightly increases the overhead.

Regarding Table~\ref{tbl:provlake_grouping}, we observe low overhead ($<$3\%) when grouping 50 messages for a task duration of 0.5 seconds, and grouping from 20 messages for a task duration of 1 second, for 1Gbit bandwidth. While for 25Kbit bandwidth, we observe high overhead ($>$43\%) for all workloads.

\renewcommand{\arraystretch}{1.3}
\begin{table}[t]
\footnotesize
\centering
\caption{Capture overhead of ProvLake and DfAnalyzer.}
\label{tbl:overhead_existing_solutions}
\begin{tabular}{crcccc}
                     & 
\textbf{\begin{tabular}[c]{@{}r@{}}overhead\\ level\end{tabular}} & \cellcolor[HTML]{D9EAD3}\textbf{\begin{tabular}[c]{@{}c@{}}low\\ $\leqslant{}$3\%\end{tabular}} & \cellcolor[HTML]{F4CCCC}\textbf{\begin{tabular}[c]{@{}c@{}}high\\ $\textgreater{}$3\%\end{tabular}} 


& \multicolumn{1}{l}{}           \\
                     &
\multicolumn{1}{l}{}                                                 & \multicolumn{1}{l}{}                                                                           & \multicolumn{1}{l}{}                                                                              & \multicolumn{1}{l}{}                                                                                    & \multicolumn{1}{l}{}           \\ \hline
                                       \textbf{
\begin{tabular}[c]{@{}c@{}}attributes\\ per task\end{tabular}
}
                                       &  
\textbf{\begin{tabular}[c]{@{}r@{}}Provenance \\ System\end{tabular}}                   & \multicolumn{4}{c}{ \textbf{Capture Overhead (\%)}}                                                                                                                                                                                                                                                                    \\ \hline
             \textbf{10}        &
ProvLake                                     & \cellcolor[HTML]{EA9999} 
\begin{tabular}[c]{@{}c@{}}56.9\%\\ $\pm$0.08\end{tabular}
& 
\cellcolor[HTML]{EA9999}

\begin{tabular}[c]{@{}c@{}}29.9\%\\ $\pm$0.29\end{tabular}
& \cellcolor[HTML]{F4CCCC}
\begin{tabular}[c]{@{}c@{}}8.56\%\\ $\pm$0.01\end{tabular}
& \cellcolor[HTML]{F4CCCC}
\begin{tabular}[c]{@{}c@{}}6.02\%\\ $\pm$0.01\end{tabular}
\\ \cline{3-6}
                                       
\textbf{10}
& DfAnalyzer                                                           & \cellcolor[HTML]{EA9999}
\begin{tabular}[c]{@{}c@{}}39.8\%\\ $\pm$0.06\end{tabular}                                                                 & \cellcolor[HTML]{F4CCCC}
\begin{tabular}[c]{@{}c@{}}21.2\%\\ $\pm$0.34\end{tabular}
& \cellcolor[HTML]{F4CCCC}
\begin{tabular}[c]{@{}c@{}}6.12\%\\ $\pm$0.07\end{tabular}

& \cellcolor[HTML]{F4CCCC}
\begin{tabular}[c]{@{}c@{}}4.26\%\\ $\pm$0.01\end{tabular}
\\ 
                                       \hline 

        \textbf{100}                               &  
ProvLake                                     & \cellcolor[HTML]{EA9999}
\begin{tabular}[c]{@{}c@{}}57.3\%\\ $\pm$0.10\end{tabular}
& \cellcolor[HTML]{EA9999}

\begin{tabular}[c]{@{}c@{}}30.1\%\\ $\pm$0.41\end{tabular}
& \cellcolor[HTML]{F4CCCC}
\begin{tabular}[c]{@{}c@{}}8.57\%\\ $\pm$0.01\end{tabular}
& \cellcolor[HTML]{F4CCCC}
\begin{tabular}[c]{@{}c@{}}6.04\%\\ $\pm$0.04\end{tabular}
\\ \cline{3-6}


\textbf{100}
& 

DfAnalyzer                                                  & \cellcolor[HTML]{EA9999}
\begin{tabular}[c]{@{}c@{}}40.5\%\\ $\pm$0.20\end{tabular}

& \cellcolor[HTML]{F4CCCC}
\begin{tabular}[c]{@{}c@{}}21.3\%\\ $\pm$0.06\end{tabular}
& \cellcolor[HTML]{F4CCCC}
\begin{tabular}[c]{@{}c@{}}6.12\%\\ $\pm$0.01\end{tabular}
& \cellcolor[HTML]{F4CCCC}
\begin{tabular}[c]{@{}c@{}}4.31\%\\ $\pm$0.01\end{tabular}
\\ \hline
                                  &
                                       
\textbf{\begin{tabular}[c]{@{}r@{}}task dur. (s)\end{tabular}}                                           & \textbf{0.5}                                                                                   & \textbf{1}                                                                                        & \textbf{3.5}                                                                                            & \textbf{5}                     \\ \hline
\end{tabular}
\end{table}

\begin{table}[t]
\footnotesize
\centering
\caption{ProvLake: impact of bandwidth and grouping strategy on the capture overhead.}
\label{tbl:provlake_grouping}

\begin{tabular}{rcc|
>{\columncolor[HTML]{EA9999}}c 
>{\columncolor[HTML]{F4CCCC}}c }
\hline
\multicolumn{1}{c}{\cellcolor[HTML]{FFFFFF}\begin{tabular}[c]{@{}c@{}}\textbf{\# of messages} \textbf{grouped}\end{tabular}} & \multicolumn{2}{c|}{\cellcolor[HTML]{FFFFFF}\textbf{Bandwidth 1Gbit}}    & \multicolumn{2}{c}{\cellcolor[HTML]{FFFFFF}\textbf{Bandwidth 25Kbit}} \\ \hline
\textbf{0}                                                                                                                & \cellcolor[HTML]{EA9999}
\begin{tabular}[c]{@{}c@{}}57.3\%\\ $\pm$0.10\end{tabular}
& \cellcolor[HTML]{EA9999}
\begin{tabular}[c]{@{}c@{}}30.1\%\\ $\pm$0.27\end{tabular}
& 
\begin{tabular}[c]{@{}c@{}}321\%\\ $\pm$1.05\end{tabular}
& \cellcolor[HTML]{EA9999}
\begin{tabular}[c]{@{}c@{}}161\%\\ $\pm$1.14\end{tabular}
\\ \cline{2-5}
\textbf{10}                                                                                                               & \cellcolor[HTML]{F4CCCC}
\begin{tabular}[c]{@{}c@{}}6.83\%\\ $\pm$0.02\end{tabular}
& \cellcolor[HTML]{F4CCCC}
\begin{tabular}[c]{@{}c@{}}3.58\%\\ $\pm$0.20\end{tabular}
& 
\begin{tabular}[c]{@{}c@{}}102.5\%\\ $\pm$3.89\end{tabular}
& 
\cellcolor[HTML]{EA9999}\begin{tabular}[c]{@{}c@{}}49.8\%\\ $\pm$2.92\end{tabular}
\\ \cline{2-5}
\textbf{20}                                                                                                              & \cellcolor[HTML]{F4CCCC}
\begin{tabular}[c]{@{}c@{}}3.87\%\\ $\pm$0.01\end{tabular}
& \cellcolor[HTML]{D9EAD3}
\begin{tabular}[c]{@{}c@{}}1.99\%\\ $\pm$0.01\end{tabular}
& 
\begin{tabular}[c]{@{}c@{}}100.8\%\\ $\pm$3.78\end{tabular}
& 
\cellcolor[HTML]{EA9999}\begin{tabular}[c]{@{}c@{}}51.16\%\\ $\pm$1.03\end{tabular}
\\ \cline{2-5}
\textbf{50}                                                                                                               & \cellcolor[HTML]{D9EAD3}
\begin{tabular}[c]{@{}c@{}}2.37\%\\ $\pm$0.01\end{tabular}
& \cellcolor[HTML]{D9EAD3}
\begin{tabular}[c]{@{}c@{}}1.24\%\\ $\pm$0.01\end{tabular}
& 
\begin{tabular}[c]{@{}c@{}}95.04\%\\ $\pm$0.10\end{tabular}
& 
\cellcolor[HTML]{EA9999}\begin{tabular}[c]{@{}c@{}}43.23\%\\ $\pm$0.28\end{tabular}
\\ \hline
\cellcolor[HTML]{FFFFFF}\textbf{\begin{tabular}[c]{@{}r@{}}task duration (s)\end{tabular}}                     & \cellcolor[HTML]{FFFFFF}\textbf{0.5}    & \cellcolor[HTML]{FFFFFF}\textbf{1}      & \cellcolor[HTML]{FFFFFF}\textbf{0.5}      & \cellcolor[HTML]{FFFFFF}\textbf{1}          \\ \hline
\end{tabular}
\end{table}

\subsection{Design-level Limitations of Existing Systems}

Table~\ref{tbl:limitations} presents the takeaways of our performance analysis and exposes the main limitations of the existing provenance systems. In summary, the evaluation shows that the existing systems present high overheads ($>$3\%) when capturing on IoT/Edge devices. 

ProvLake and DfAnalyzer rely on HTTP over TCP, instead of IoT-based messaging and transmission protocols such as MQTT~\cite{mqtt}, CoAP~\cite{coap}, AMQP~\cite{amqp}, UDP~\cite{udp}, RPL~\cite{rpl}, to cite a few. In resource-constrained devices, they make a relevant impact on performance, resource usage, and power consumption, as explored by existing works~\cite{morabito2018evaluating, wukkadada2018comparison, gemirter2021comparative}.

The experiment results reinforce the need for capture approaches tailored to the constraints imposed by IoT devices. In addition, simplified data models to represent the provenance data help to reduce overheads.

\begin{table}[t]
\small
\centering
\caption{Limitations of existing provenance systems.}
\label{tbl:limitations}
\begin{tabular}{m{1.4cm}m{6.1cm}}
\hline
\textbf{\begin{tabular}[c]{@{}r@{}}System\end{tabular}}
& \textbf{Limitation}                                                                                                             \\ \hline
\rowcolor[HTML]{D0E0E3} 
{\color[HTML]{434343} DfAnalyzer} & {\color[HTML]{434343} Presents \textbf{high} ($\textgreater$3\%) \textbf{capture overhead} for all synthetic workloads.}                                                                                                              \\
ProvLake                          & {\color[HTML]{434343} Presents \textbf{high} ($\textgreater$3\%) \textbf{overhead} for all workloads. However, ProvLake allows \textbf{grouping captured data} to reduce transmission frequency, enabling lower overhead, but it \textbf{still suffers high overhead} in \textbf{low bandwidth} networks. 
} \\
\rowcolor[HTML]{D0E0E3} 
{\color[HTML]{434343} PROV-IO}    & {\color[HTML]{434343}Does not send the captured data over the network to another machine hosting the provenance system. Instead, it periodically \textbf{dumps the in-memory provenance graph to disk}. This approach is \textbf{not suitable} for IoT/Edge devices.}                                                                                                                                                                            \\
{\color[HTML]{434343} Komadu}     & {\color[HTML]{434343} Komadu does not follow a clear separation between a client library and a backend provenance server. Therefore, the \textbf{capture} and the \textbf{processing} of the captured information \textbf{run in the same machine}. This approach is \textbf{not suitable} for capturing on IoT/Edge devices.}                                                                                                                                                                                         \\ \hline
\end{tabular}
\end{table}

\section{ProvLight Design}
\label{sec:provlight}

\begin{table*}[t]
\footnotesize
\centering
\caption{The ProvLight provenance data exchange model follows PROV-DM.}
\label{tbl:provlight-w3c}
\begin{tabular}{lllll}
\hline
{\color[HTML]{434343} \textbf{\begin{tabular}[c]{@{}l@{}}PROV-DM \\Type\end{tabular}}} & {\color[HTML]{434343} \textbf{\begin{tabular}[c]{@{}l@{}}ProvLight \\Class\end{tabular}}} & {\color[HTML]{434343} \textbf{\begin{tabular}[c]{@{}l@{}}ProvLight \\ Class Attributes\end{tabular}}}                                                         & {\color[HTML]{434343} \textbf{\begin{tabular}[c]{@{}l@{}}ProvLight Attribute Description \\ and PROV-DM Relationships\end{tabular}}}                                                                                                                                                                                                                                                                                         & {\color[HTML]{434343} \textbf{\begin{tabular}[c]{@{}l@{}}ProvLight\\ Class Description\end{tabular}}}   \\ \hline
\rowcolor[HTML]{D0E0E3} 
{\color[HTML]{434343} Agent}                  & {\color[HTML]{434343} Workflow}                 & {\color[HTML]{434343} 
\begin{tabular}[c]{@{}l@{}}
id
\end{tabular}}                                                                                      & {\color[HTML]{434343} 
\begin{tabular}[c]{@{}l@{}}
Workflow id.
\end{tabular}}                                                                                                                                                                                                                                                                                                                             & {\color[HTML]{434343} \begin{tabular}[c]{@{}l@{}}Refers to application workflows.\end{tabular}}                                                \\
{\color[HTML]{434343} Activity}               & {\color[HTML]{434343} Task}                     & {\color[HTML]{434343} 

\begin{tabular}[c]{@{}l@{}}
id\\ 
workflow\\  
dependencies\\  
data\\ 
time\\
status
\end{tabular}} 
& 
{\color[HTML]{434343} \begin{tabular}[c]{@{}l@{}}
Task id.\\ 
Links tasks with the workflow they belong to (\textit{wasAssociatedWith}).\\ 
Dependencies between tasks (\textit{wasInformedBy}).\\ 
Data used (\textit{used}) and generated (\textit{wasGeneratedBy}) by a task.\\
Task start and end time.\\ 
Task status: running or finished.
\end{tabular}} 
& {\color[HTML]{434343} 
\begin{tabular}[c]{@{}l@{}}Represents the processing \\steps of tasks (and their \\dependencies) that compose \\workflows.\end{tabular}
} \\
\rowcolor[HTML]{D0E0E3} 
{\color[HTML]{434343} Entity}                 & {\color[HTML]{434343} Data}           & {\color[HTML]{434343} 
\begin{tabular}[c]{@{}l@{}}
id\\ 
workflow\_id\\ 
derivations\\ 
attributes
\end{tabular}}                                                                            & {\color[HTML]{434343} 
\begin{tabular}[c]{@{}l@{}}
Data id.\\ 
Links data with the workflow they belong to (\textit{wasAttributedTo}).\\ 
Links chained data (\textit{wasDerivedFrom}).\\ 
Data attributes and values. 
\end{tabular}}                                                                                                                                                                                                                                                                                            & {\color[HTML]{434343} \begin{tabular}[c]{@{}l@{}}Represents data\\ derivations along\\ the workflow execution.\end{tabular}}                        \\ \hline
\end{tabular}
\end{table*}

This section introduces ProvLight, a tool~\cite{git-provlight} for the efficient provenance data capture of Edge-to-Cloud workflows. ProvLight is designed to capture provenance in IoT/Edge devices with low overhead in terms of capture time, CPU and memory usage, network usage, and power consumption. 

Subsection~\ref{subsec:provlight_model} presents the ProvLight provenance model. Next, the architectural details are given in Subsection~\ref{subsec:provlight_arch}, while Subsection~\ref{subsec:provlight_impl} describes its implementation.

\subsection{Data Exchange Model} 
\label{subsec:provlight_model}

ProvLight provenance data exchange model follows the W3C PROV-DM~\cite{belhajjame2013prov} recommendation. The goal is to have a data exchange schema (domain-agnostic PROV modeling) for capturing data in the IoT/Edge and making sure these captured data are compatible with W3C PROV-based workflow provenance systems, such as ProvLake, DfAnalyzer, PROV-IO, among many others. Table~\ref{tbl:provlight-w3c} describes ProvLight classes and their relationships and maps them to PROV-DM core elements.

The main classes of our model are \emph{Workflow}, \emph{Task}, and \emph{Data}. These classes are derived from the \emph{Agent}, \emph{Activity}, and \emph{Entity} PROV-DM types, respectively. ProvLight classes aim to provide a simplified abstraction allowing users to track workflow (\emph{Workflow} class), input and output parameters (\emph{Data} class), and processing history (\emph{Task} class).

The \emph{Workflow} class may be used to refer to the application workflow (\emph{e.g.,} Federated Learning training). The \emph{Task} class refers to the tasks executed in the workflow (\emph{e.g.,} each epoch or model update of the model training). Finally, the \emph{Data} class represents the input data attributes and values (\emph{e.g.,} hyperparameters of the learning algorithm) or the output attributes (\emph{e.g.,} training time and loss of each epoch).

To represent PROV-DM relationships, we use the \emph{id} attribute of each class. We link the \emph{Task} and \emph{Data} classes with the workflow they belong to (\emph{wasAssociatedWith} and \emph{wasAttributedTo}, respectively). The links between a \emph{Task} and its respective \emph{Data} inputs and the generated outputs are represented by the \emph{used} and \emph{wasGeneratedBy} relationships, respectively. The \emph{dependencies} attribute in the \emph{Task} class links tasks (\emph{wasInformedBy}) with dependencies (\emph{e.g.,} task \emph{B} starts after task \emph{A} ends). Finally, the \emph{derivations} attribute in the \emph{Data} class links (\emph{wasDerivedFrom}) chained data (\emph{e.g.,} data $D_A$ was used in task \emph{A} to generate data $D_B$).

Defining such relations aims to provide users with the data processing history: \emph{Where} did the data come from? \emph{How} was the data transformed? and \emph{Who} acted upon it? For instance, capturing provenance data of Federated Learning model training workflows may help users to interpret results. Tracking model training at runtime and fine-tuning hyperparameters is helpful, especially when the training process takes a long time.

\begin{figure}[t]
  \centering
  \includegraphics[width=\linewidth]{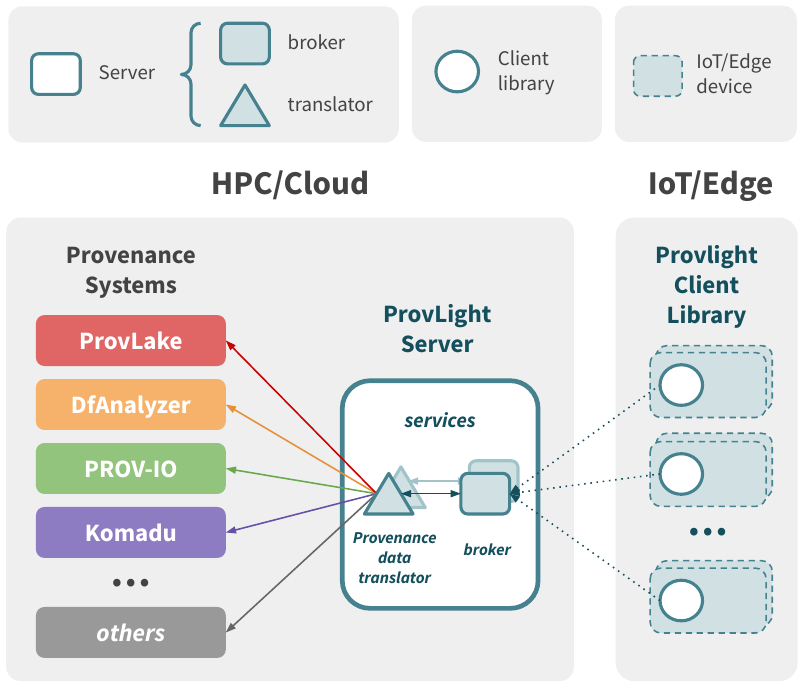}
  \caption{ProvLight Architecture.}
  \label{fig:provlight-arch}
\end{figure}

\begin{table}[t]
\scriptsize
\centering
\caption{How does ProvLight differ from state-of-the-art systems in terms of data capture?}
\label{tbl:provlight-details}
\begin{tabular}{r>{\columncolor[HTML]{D0E0E3}}c cc}
\hline
\multicolumn{1}{l}{}                                                                                & {\color[HTML]{434343} \textbf{ProvLight}}                                                                                                         & {\color[HTML]{434343} \textbf{DfAnalyzer}}                                         & {\color[HTML]{434343} \textbf{ProvLake}}                                                 \\ \hline
{\color[HTML]{434343} \textbf{\begin{tabular}[c]{@{}r@{}}application layer\\ protocol\end{tabular}}}                                          & {\color[HTML]{434343} \begin{tabular}[c]{@{}c@{}}MQTT-SN\\ \textit{QoS 2: Exactly once}\end{tabular}}                                                     & {\color[HTML]{434343} HTTP 1.1}                                                    & {\color[HTML]{434343} HTTP 1.1}                                                          \\ \hline
{\color[HTML]{434343} \textbf{\begin{tabular}[c]{@{}r@{}}transport layer\\ protocol\end{tabular}}}                                            & {\color[HTML]{434343} UDP}                                                                                                                        & {\color[HTML]{434343} TCP}                                                         & {\color[HTML]{434343} TCP}                                                               \\ \hline
{\color[HTML]{434343} \textbf{\begin{tabular}[c]{@{}r@{}}Communication\\ model\end{tabular}}}      & {\color[HTML]{434343} \begin{tabular}[c]{@{}c@{}}Publish/\\ Subscribe\end{tabular}}                                                               & {\color[HTML]{434343} \begin{tabular}[c]{@{}c@{}}Request/\\ Response\end{tabular}} & {\color[HTML]{434343} \begin{tabular}[c]{@{}c@{}}Request/\\ Response\end{tabular}}       \\ \hline
{\color[HTML]{434343} \textbf{Server side}}                                                         & {\color[HTML]{434343} \begin{tabular}[c]{@{}r@{}}MQTT-SN Broker\end{tabular}}                                                                                                             & {\color[HTML]{434343} \begin{tabular}[c]{@{}r@{}}HTTP\\ Server\end{tabular}}                                                 & {\color[HTML]{434343} \begin{tabular}[c]{@{}r@{}}HTTP\\ Server\end{tabular}}                                                       \\ \hline
{\color[HTML]{434343} \textbf{\begin{tabular}[c]{@{}r@{}}Client side\\ features\end{tabular}}} & {\color[HTML]{434343} \begin{tabular}[c]{@{}c@{}}provenance data\\ representation \& \\ payload compression \&\\ grouping data captured\end{tabular}} & {\color[HTML]{434343} N/A}                                                         & {\color[HTML]{434343} \begin{tabular}[c]{@{}c@{}}grouping data \\ captured\end{tabular}} \\ \hline

{\color[HTML]{434343} \textbf{\begin{tabular}[c]{@{}r@{}}Provenance data\\ model\end{tabular}}} & {\color[HTML]{434343} PROV-DM} & {\color[HTML]{434343} PROV-DM} & {\color[HTML]{434343} PROV-DM} \\ \hline

{\color[HTML]{434343} \textbf{\begin{tabular}[c]{@{}r@{}}Capture library\\ language\end{tabular}}} & {\color[HTML]{434343} Python} & {\color[HTML]{434343} Python, C++} & {\color[HTML]{434343} Python} \\ \hline

\end{tabular}
\end{table}

\subsection{Architecture}
\label{subsec:provlight_arch}

Figure~\ref{fig:provlight-arch} presents the ProvLight architecture. It follows a \textit{client/server} model where the \textit{server} receives the captured data from \textit{clients} and then translates it and sends it to provenance systems. We highlight that ProvLight may integrate with existing provenance systems like DfAnalyzer, ProvLake, and PROV-IO, among others (\emph{e.g.,} through their APIs and ProvLight data translator), as a solution for capturing data of workflows running on IoT/Edge devices, as illustrated in Figure~\ref{fig:provlight-arch}. Table~\ref{tbl:provlight-details} summarizes how the ProvLight architecture design differs from the systems analyzed in Section~\ref{sec:soalimitations}.

This integration may be achieved by using:

\subsubsection{\textbf{Server}} The ProvLight \emph{server} is composed of a \emph{broker} and a \emph{provenance data translator}. Both may be parallelized to scale the data capture for scenarios with various IoT/Edge devices. We describe the main roles of each one.

\textbf{\emph{(i) Broker:}} refers to an MQTT-SN broker (MQTT for Sensor Networks~\cite{stanford2013mqtt}). During workflow execution, \emph{clients} subscribe to the \emph{broker} and then start to transmit the captured data. Next, this data is forwarded to the \emph{provenance data translator}, which is subscribed to the \emph{broker}. 

\textbf{\emph{(ii) Provenance Data Translator:}} translates the captured data to the respective format used by the provenance system. The \emph{provenance data translator} may be extended, by users, to translate to a particular data model of a provenance system. After translating, it sends the data to the provenance system service (\emph{e.g.,} typically available at an \emph{ip:port}). It allows seamless integration with existing systems.

\subsubsection{\textbf{Client}} The ProvLight \emph{client} aims to efficiently capture provenance data on resource-limited devices. ProvLight provides a client library that follows the W3C PROV-DM provenance model (as presented in Table~\ref{tbl:provlight-w3c}). This library allows users to instrument their workflow code to decide what data to capture. A \emph{client} is configured to transmit, at runtime, the captured data to the remote \emph{broker} (\emph{e.g.,} \emph{ip:port}). This allows users to track workflow execution at runtime (\emph{e.g.,} started and finished tasks, input and output data, \emph{etc.}) through provenance systems supporting data ingestion at runtime.

\begin{figure}[t]
    \lstset{aboveskip=0pt,belowskip=0pt}
\begin{lstlisting}[language=Python, escapechar=|, caption=ProvLight: user-defined provenance capture., label=lis-udp, 
    basicstyle=\scriptsize
    ]
from provlight import Workflow, Task, Data
attributes = 100
chained_transformations = 5
number_of_tasks = 100
# Application Workflow
workflow = Workflow(1)|\label{line:wf}|
workflow.begin()|\label{line:wf-b}|
# Tasks and data derivations
data_id = 0
previous_task = []
in_data = {'in': [1 for _ in range(attributes)]}
out_data = {'out': [2 for _ in range(attributes)]}
for transf_id in range(chained_transformations):
    for task_id in range(int(number_of_tasks/chained_transformations)):
        data_id += 1
        task = Task(transf_id-task_id, workflow, transf_id, dependencies=previous_task)|\label{line:tk}|
        data_in= Data(in_{data_id}, workflow.id, in_data)|\label{line:dd-in}|
        task.begin([data_in])|\label{line:tk-b}|
        #### ADD YOUR TASK HERE|\label{line:tk-run}| ####
        data_out= Data(out_{data_id}, workflow.id,out_data)|\label{line:dd-out}|
        task.end([data_out])|\label{line:tk-e}|
        previous_task = [task.id]
workflow.end()|\label{line:wf-e}|
\end{lstlisting}
\end{figure}

\subsection{Implementation}
\label{subsec:provlight_impl}

\subsubsection{\textbf{Server}} The \textbf{\emph{Broker}} is implemented based on the Eclipse RSMB server~\cite{code-rsmb} (Really Small Message Broker). RSMB builds on top of Mosquitto~\cite{code-mosquitto} and implements the MQTT-SN protocol.

The \textbf{\emph{Provenance Data Translator}} is a Python service that may be extended to translate captured data (from the ProvLight data format) to a particular provenance system (\emph{e.g.,} DfAnalyzer, ProvLake, Komadu, \emph{etc.}). In our repository~\cite{git-provlight}, we provide an implementation showing how to translate from the ProvLight data format to DfAnalyzer. Such translation is possible since the aforementioned systems follow the W3C PROV-DM provenance model. For the \emph{translator-to-broker} communication, we use the MQTT-SN Python client library~\cite{code-mqttsn} based on Eclipse RSMB. Finally, for the \emph{translator-to-provenance-system} communication, users are free to use any Python library compatible with the provenance system (\emph{e.g.,} Requests~\cite{code-requests}).


\subsubsection{\textbf{Client}} The ProvLight client library is implemented in Python and provides a series of features targeting resource-limited IoT/Edge devices:

\begin{itemize}
    \item \emph{provenance data representation:} simplified classes for provenance modeling that allow users to represent workflows, data derivations (\emph{e.g.,} input/output data from tasks) and tasks (\emph{e.g.,} status, dependencies, data derivations);

    \item \emph{payload compression:} compresses the bytes in captured data before transmitting over the network; and

    \item \emph{data capture grouping:} allow users to optionally group data just from ended tasks, so users may still track at workflow runtime the tasks that have already started.
\end{itemize}

As shown later in the evaluation section, grouping and compressing captured data help reduce capture time overhead, especially in IoT/Edge devices.

\textbf{\textit{How to capture provenance data from the workflows?}} Listing~\ref{lis-udp} illustrates an example of application code instrumentation with the ProvLight library highlighted in blue color. Lines~\ref{line:wf},~\ref{line:wf-b}, and~\ref{line:wf-e} instantiate the workflow, start, and finalize it, respectively. Line~\ref{line:tk} instantiates a task, linking it to the \emph{workflow}, input data \emph{derivation}, and dependent task. Lines~\ref{line:tk-b} and~\ref{line:tk-e} capture data from the initialization and finalization of the task. Before starting a task, line~\ref{line:dd-in} instantiates \emph{Data} and adds it as input data (line~\ref{line:tk-b}) to the task. Following the same logic, line~\ref{line:dd-out} instantiates and adds the output data from the task. We highlight that the \emph{begin()} and \emph{end()} methods of \emph{Workflow} and \emph{Task} transmit the captured data over the network to the \emph{broker}. Finally, line~\ref{line:tk-run} is where the workflow task runs.

\section{Provenance Capture of Edge-to-Cloud Workflows}
\label{sec:provlight_e2clab_extension}


This section presents the integration of ProvLight as a key system in the E2Clab~\cite{rosendo:hal-02916032} framework for reproducible experimentation across the Edge-to-Cloud Continuum. This integration allows users to capture end-to-end provenance data of Edge-to-Cloud workflows.  
Figure~\ref{fig:e2clab-prov} shows the extended E2Clab architecture with the new components in the red color.

\subsection{Provenance Manager} 
\begin{figure}[t]
  \centering
  \includegraphics[width=\linewidth]{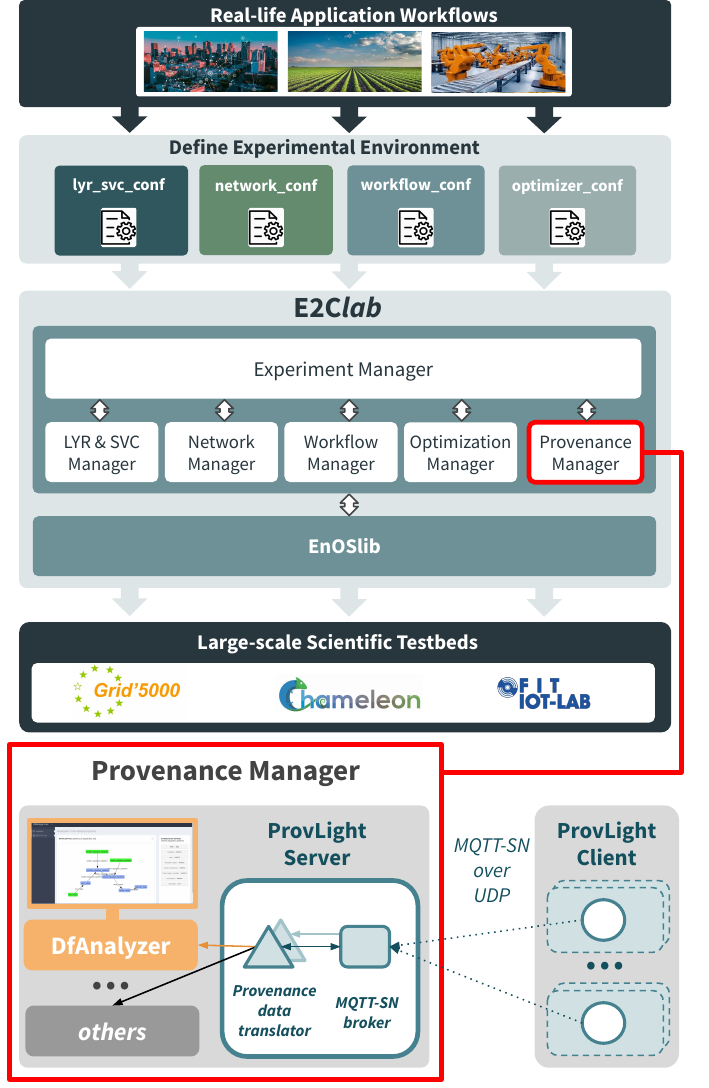}
  \caption{Extended E2Clab: Provenance Data Manager.}
  \label{fig:e2clab-prov}
\end{figure}

We design a new manager named \emph{Provenance Manager}. Figure~\ref{fig:e2clab-prov} illustrates the integrated view of the two main elements that compose the Provenance Manager:

\textbf{\emph{(i) ProvLight:}} to efficiently capture provenance data of workflows running on IoT devices. It also allows users to capture provenance in Cloud/HPC environments. ProvLight translates the captured data to the DfAnalyzer data model. 

\textbf{\emph{(ii) DfAnalyzer:}} to store and query provenance captured by \emph{ProvLight} during workflow runtime (\emph{e.g.,} compare provenance of multiple workflow evaluations to understand how they impact on performance). Furthermore, it allows users to visualize dataflow specifications (\emph{i.e.,} data attributes of each dataset).

In addition to the characteristics of the provenance systems analyzed in Table~\ref{tbl:limitations}, and due to ProvLake being proprietary within IBM, while DfAnalyzer is open source~\cite{dfa-code}, in this work we decide to use DfAnalyzer. As the data capture component of DfAnalyzer presents high overhead, we just use its data analysis and storage components. Finally, the \emph{Provenance Manager} could replace DfAnalyzer with other provenance systems (\emph{e.g.,} PROV-IO, Komadu, \emph{etc.}). It requires extending ProvLight to translate the provenance data to the data model of the provenance system and using their APIs.

\lstset{
  emph={environment, envir, name, cluster, layers, services, quantity, qtd, roles, env_name, iotlab, g5k, archi, arch, job_name, job_type, walltime, provenance},
  emphstyle={\color{codeorange}}%
}

\subsection{Provenance Capture}

Through the E2Clab framework, users may easily enable provenance data capture across the Edge-to-Cloud continuum through simple configuration files, as illustrated in Listing~\ref{lis-e2c-prov}. Listing~\ref{lis-e2c-prov} refers to the E2Clab \emph{layers\_services.yaml} configuration file used to setup the experimental environment (\emph{e.g.,} testbeds, services that compose workflows, \emph{etc.}). Lines~\ref{line:g5k} and~\ref{line:iotlab} request resources from Grid'5000 and FIT IoT LAB testbeds, respectively. Line~\ref{line:server} requests a single server (\emph{e.g.,} Federated Learning server) on the Cloud layer; while line~\ref{line:client} requests 64 clients (\emph{e.g.,} to train the model with their local data) on the Edge layer. Finally, line~\ref{line:provmanager} setups the provenance data capture (the \emph{ProvenanceManager} service). After that, users may instrument their application code to capture data, as presented in Listing~\ref{lis-udp}.



The \emph{ProvenanceManager} service starts a Docker~\cite{docker-cont} container with the DfAnalyzer provenance system and a ProvLight container allowing clients to send their provenance data. DfAnalyze exhibits at workflow runtime the captured data on its Web interface. The \emph{ProvenanceManager} service may be easily plugged into other provenance systems by just using their  Docker images and extending the provenance data translator.

\begin{figure}[t]
    \lstset{aboveskip=0pt,belowskip=0pt}
\begin{lstlisting}[language=Python, escapechar=|, caption=E2Clab: provenance of Edge-to-Cloud workflows., label=lis-e2c-prov,
    basicstyle=\scriptsize
    ]
environment:
  g5k:|\label{line:g5k}| cluster: gros
  iotlab:|\label{line:iotlab}| cluster: grenoble
provenance:|\label{line:provmanager}| ProvenanceManager
layers:
- name: cloud
  services:
  - name: Server|\label{line:server}|, environment: g5k, qtd: 1
- name: edge
  services:
  - name: Client|\label{line:client}|, environment: iotlab, arch: a8, qtd: 64
\end{lstlisting}
\end{figure}

\section{Evaluation}
\label{sec:evaluation}

\begin{figure}[t]
  \centering
  \includegraphics[width=\linewidth]{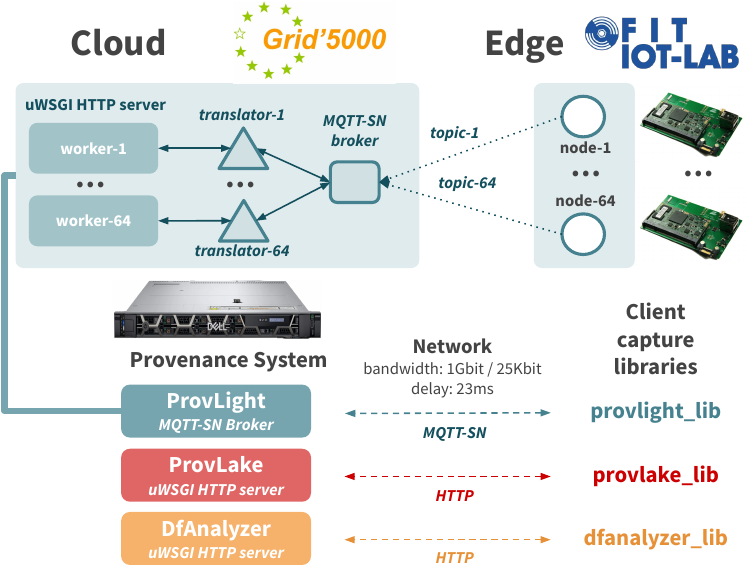}
  \caption{Experimental setup.}
  \label{fig:exp_setup}
\end{figure}

We aim to answer the following research questions: \emph{\textbf{how does ProvLight perform in IoT/Edge devices?} while initially targeting resource-constrained Edge devices, \textbf{can ProvLight be efficiently used also in the Cloud?}} We answer these questions in subsections A---D and E, respectively, by comparing ProvLight against ProvLake and DfAnalyzer.

The \emph{main performance metric} is the capture overhead  in terms of: \emph{(i)} data capture time; \emph{(ii)} CPU and memory usage; \emph{(iii)} network usage; and \emph{(iv)} power consumption. \textbf{The experimental setup is the same as presented in Subsection~\ref{subsec:setup}}, with synthetic workloads generated based on the Federated Learning use case. The deployment is shown in Figure~\ref{fig:exp_setup}. Results in Figure~\ref{fig:overhead_analysis} are the mean of 10 runs with their 95\% confidence interval.


\subsection{Capture Time Overhead} 

Table~\ref{tbl:overhead_comparison} presents the capture time overhead comparison for the 8 synthetic workloads. \textbf{In summary, ProvLight presents low capture overhead ($<$3\%) for all workloads analyzed.} Regarding tasks with a duration of 3.5 seconds or more, the capture overhead of ProvLight is below 0.5\%. Varying the number of attributes per task from 10 to 100 does not significantly increase the capture time. We highlight that \textbf{ProvLight is about 37x and 26x faster than ProvLake and DfAnalyzer, respectively.}

Similarly to Table~\ref{tbl:provlake_grouping}, Table~\ref{tbl:impact_of_bandwidth} zooms our analysis in order to understand the impact of bandwidth variations and the grouping strategy on the data capture time. Results show that, differently from ProvLake, \textbf{ProvLight presents low capture time overhead in low-bandwidth scenarios for task durations of 0.5 and 1 second.} We highlight that, especially in low-bandwidth scenarios (25Kbit), the ProvLight grouping strategy presents low overhead ($<$2\%), while ProvLake presents high overhead ($>$43\%), see Table~\ref{tbl:provlake_grouping}.

\renewcommand{\arraystretch}{1.3}
\begin{table}[t]
\footnotesize
\centering
\caption{ProvLight: capture overhead in IoT/Edge devices. Refer to Table~\ref{tbl:overhead_existing_solutions} to compare with DfAnalyzer and ProvLake.}
\label{tbl:overhead_comparison}
\begin{tabular}{rcccc}
                                       \hline
\textbf{\begin{tabular}[c]{@{}r@{}}Attributes per task\end{tabular}}                   & \multicolumn{4}{c}{ \textbf{Capture Overhead (\%)}}                                                                                                                                                                                                                                                                    \\ \hline

                    \textbf{10}                                                                                     & \cellcolor[HTML]{D9EAD3}
\begin{tabular}[c]{@{}c@{}}1.45\%\\ $\pm$0.01\end{tabular}
& \cellcolor[HTML]{D9EAD3}
\begin{tabular}[c]{@{}c@{}}1.02\%\\ $\pm$0.01\end{tabular}
& \cellcolor[HTML]{D9EAD3}
\begin{tabular}[c]{@{}c@{}}0.31\%\\ $\pm$0.01\end{tabular}
& \cellcolor[HTML]{D9EAD3}
\begin{tabular}[c]{@{}c@{}}0.23\%\\ $\pm$0.01\end{tabular}

\\ \hline

                                 \textbf{100}                                                                            & \cellcolor[HTML]{D9EAD3}
\begin{tabular}[c]{@{}c@{}}1.54\%\\ $\pm$0.01\end{tabular}
& \cellcolor[HTML]{D9EAD3}
\begin{tabular}[c]{@{}c@{}}1.11\%\\ $\pm$0.01\end{tabular}
& \cellcolor[HTML]{D9EAD3}
\begin{tabular}[c]{@{}c@{}}0.37\%\\ $\pm$0.01\end{tabular}
& \cellcolor[HTML]{D9EAD3}
\begin{tabular}[c]{@{}c@{}}0.29\%\\ $\pm$0.01\end{tabular}
\\
\hline

\textbf{\begin{tabular}[c]{@{}r@{}}task duration (s)\end{tabular}}                                           & \textbf{0.5}                                                                                   & \textbf{1}                                                                                        & \textbf{3.5}                                                                                            & \textbf{5}                     \\ \hline
\end{tabular}
\end{table}

\begin{table}[t]
\footnotesize
\centering
\caption{ProvLight: impact of bandwidth and grouping strategy on the capture overhead. Refer to Table~\ref{tbl:provlake_grouping} to compare with ProvLake.}
\label{tbl:impact_of_bandwidth}

\begin{tabular}{rcc|
>{\columncolor[HTML]{EA9999}}c 
>{\columncolor[HTML]{F4CCCC}}c }
\hline
\multicolumn{1}{c}{\cellcolor[HTML]{FFFFFF}\begin{tabular}[c]{@{}c@{}}\textbf{\# of messages} \textbf{grouped}\end{tabular}} & \multicolumn{2}{c|}{\cellcolor[HTML]{FFFFFF}\textbf{Bandwidth 1Gbit}}    & \multicolumn{2}{c}{\cellcolor[HTML]{FFFFFF}\textbf{Bandwidth 25Kbit}} 
\\ \hline

\textbf{0}                                                                               & \cellcolor[HTML]{D9EAD3}
\begin{tabular}[c]{@{}c@{}}1.54\%\\ $\pm$0.01\end{tabular}
& \cellcolor[HTML]{D9EAD3}
\begin{tabular}[c]{@{}c@{}}1.10\%\\ $\pm$0.01\end{tabular}
& \cellcolor[HTML]{D9EAD3}
\begin{tabular}[c]{@{}c@{}}1.56\%\\ $\pm$0.01\end{tabular}
& \cellcolor[HTML]{D9EAD3}
\begin{tabular}[c]{@{}c@{}}1.04\%\\ $\pm$0.01\end{tabular}
\\ \cline{2-5}
\textbf{10}                                                                              

& \cellcolor[HTML]{D9EAD3}
\begin{tabular}[c]{@{}c@{}}1.37\%\\ $\pm$0.01\end{tabular}
& \cellcolor[HTML]{D9EAD3}
\begin{tabular}[c]{@{}c@{}}0.75\%\\ $\pm$0.01\end{tabular}
& \cellcolor[HTML]{D9EAD3}
\begin{tabular}[c]{@{}c@{}}1.37\%\\ $\pm$0.01\end{tabular}
& \cellcolor[HTML]{D9EAD3}
\begin{tabular}[c]{@{}c@{}}0.74\%\\ $\pm$0.01\end{tabular}
\\ \cline{2-5}
\textbf{20}                                                                              

& \cellcolor[HTML]{D9EAD3}
\begin{tabular}[c]{@{}c@{}}1.32\%\\ $\pm$0.01\end{tabular}
& \cellcolor[HTML]{D9EAD3}
\begin{tabular}[c]{@{}c@{}}0.72\%\\ $\pm$0.01\end{tabular}
& \cellcolor[HTML]{D9EAD3}
\begin{tabular}[c]{@{}c@{}}1.34\%\\ $\pm$0.01\end{tabular}
& \cellcolor[HTML]{D9EAD3}
\begin{tabular}[c]{@{}c@{}}0.73\%\\ $\pm$0.01\end{tabular}
\\ \cline{2-5}
\textbf{50}                                                                              

& \cellcolor[HTML]{D9EAD3}
\begin{tabular}[c]{@{}c@{}}1.31\%\\ $\pm$0.01\end{tabular}
& \cellcolor[HTML]{D9EAD3}
\begin{tabular}[c]{@{}c@{}}0.72\%\\ $\pm$0.01\end{tabular}
& \cellcolor[HTML]{D9EAD3}
\begin{tabular}[c]{@{}c@{}}1.31\%\\ $\pm$0.01\end{tabular}
& \cellcolor[HTML]{D9EAD3}
\begin{tabular}[c]{@{}c@{}}0.72\%\\ $\pm$0.01\end{tabular}

\\ \hline
\cellcolor[HTML]{FFFFFF}\textbf{\begin{tabular}[c]{@{}r@{}}task duration (s)\end{tabular}}                     & \cellcolor[HTML]{FFFFFF}\textbf{0.5}    & \cellcolor[HTML]{FFFFFF}\textbf{1}      & \cellcolor[HTML]{FFFFFF}\textbf{0.5}      & \cellcolor[HTML]{FFFFFF}\textbf{1}          \\ \hline
\end{tabular}
\end{table}

\textbf{\textit{Scalability analysis.}} Table~\ref{tbl:overhead_scaling} presents the capture time overhead of ProvLight when scaling the number of IoT/Edge devices and considering 100 tasks of 0.5s each and 100 attributes per task. We scale the scenario with 8, 16, 32, and 64 devices capturing provenance data in parallel and sending the data to the cloud server. As illustrated in Figure~\ref{fig:exp_setup}, each client sends its data to its respective topic in the \emph{Broker} and we parallelized the number of \emph{translators} accordingly. Lastly, provenance systems (\emph{i.e.,} DfAnalyzer in our case) can handle  parallel requests and store the provenance data in a database system (\emph{e.g.,} MonetDB~\cite{boncz2005monetdb} used in DfAnalyzer). Results show that \textbf{by scaling up to 64 devices, the capture overhead is low ($<$3\%)} and does not significantly impact the capture time. This is expected because devices (clients) asynchronously publish their messages to their respective topics in the \emph{MQTT-SN Broker}. For 8 and 64 devices, the capture time overhead is 1.54\% and 1.57\%, respectively.


\subsection{CPU and Memory Overhead} 

Figures~\ref{fig:cpu_overhead} and~\ref{fig:memory_overhead} present the CPU and memory overhead for capturing provenance data with ProvLake, DfAnalyzer, and ProvLight (from left to right). Regarding the CPU overhead, \textbf{ProvLight uses 7x and 5x less CPU than ProvLake and DfAnalyzer}, respectively. Capturing with ProvLight, the CPU overhead is low ($<$3\%), and CPU usage varies between 1.7\% and 2\%. Regarding the memory overhead, ProvLight memory usage is $<$4\%. It uses about 2x less memory than ProvLake and DfAnalyzer.

\renewcommand{\arraystretch}{1.3}
\begin{table}[t]
\footnotesize
\centering
\caption{ProvLight scalability analysis.}
\label{tbl:overhead_scaling}

\begin{tabular}{m{0.8cm}rcccc}
                                       \hline
                                                                       &
\textbf{\begin{tabular}[c]{@{}r@{}}System\end{tabular}}                   & \multicolumn{4}{c}{ \textbf{Capture Overhead (\%)}}                                                                                                  \\ \hline

& ProvLight                                                                                     & \cellcolor[HTML]{D9EAD3}
\begin{tabular}[c]{@{}c@{}}1.54\%\\ $\pm$0.01\end{tabular}
& \cellcolor[HTML]{D9EAD3}
\begin{tabular}[c]{@{}c@{}}1.54\%\\ $\pm$0.01\end{tabular}
& \cellcolor[HTML]{D9EAD3}
\begin{tabular}[c]{@{}c@{}}1.56\%\\ $\pm$0.01\end{tabular}
& \cellcolor[HTML]{D9EAD3}
\begin{tabular}[c]{@{}c@{}}1.57\%\\ $\pm$0.02\end{tabular}
\\ \hline
                                                                     &
\textbf{\begin{tabular}[c]{@{}r@{}}\# of devices\end{tabular}} 
& \textbf{8} & \textbf{16} & \textbf{32} & \textbf{64}
\\ \hline
\end{tabular}
\end{table}

\renewcommand{\arraystretch}{1.3}
\begin{table}[t]
\footnotesize
\centering
\caption{Capture overhead in Cloud servers.}
\label{tbl:overhead_comparison_cloud}

\begin{tabular}{m{0.8cm}rcccc}
                                       \hline
                                                                       &
\textbf{\begin{tabular}[c]{@{}r@{}}System\end{tabular}}                   & \multicolumn{4}{c}{ \textbf{Capture Overhead (\%)}}                                                                                                  \\ \hline
                                                                       &  
ProvLake                                     & \cellcolor[HTML]{D9EAD3}
\begin{tabular}[c]{@{}c@{}}1.71\%\\ $\pm$0.03\end{tabular}
& \cellcolor[HTML]{D9EAD3}
\begin{tabular}[c]{@{}c@{}}0.92\%\\ $\pm$0.01\end{tabular}
& \cellcolor[HTML]{D9EAD3}
\begin{tabular}[c]{@{}c@{}}0.34\%\\ $\pm$0.01\end{tabular}
& \cellcolor[HTML]{D9EAD3}
\begin{tabular}[c]{@{}c@{}}0.26\%\\ $\pm$0.01\end{tabular}
\\ \cline{3-6}
                                      
\parbox[t]{12mm}{\multirow{-2}{*}{\textbf{\begin{tabular}[c]{@{}c@{}}100\\ attributes\\ per task\end{tabular}}}}  
                                       
&  DfAnalyzer                                                           & \cellcolor[HTML]{D9EAD3}
\begin{tabular}[c]{@{}c@{}}1.17\%\\ $\pm$0.02\end{tabular}
& \cellcolor[HTML]{D9EAD3}
\begin{tabular}[c]{@{}c@{}}0.63\%\\ $\pm$0.01\end{tabular}
& \cellcolor[HTML]{D9EAD3}
\begin{tabular}[c]{@{}c@{}}0.25\%\\ $\pm$0.01\end{tabular}
& \cellcolor[HTML]{D9EAD3}
\begin{tabular}[c]{@{}c@{}}0.21\%\\ $\pm$0.01\end{tabular}
\\ \cline{3-6}
& ProvLight                                                                                     & \cellcolor[HTML]{D9EAD3}
\begin{tabular}[c]{@{}c@{}}0.24\%\\ $\pm$0.01\end{tabular}
& \cellcolor[HTML]{D9EAD3}
\begin{tabular}[c]{@{}c@{}}0.17\%\\ $\pm$0.01\end{tabular}
& \cellcolor[HTML]{D9EAD3}
\begin{tabular}[c]{@{}c@{}}0.12\%\\ $\pm$0.01\end{tabular}
& \cellcolor[HTML]{D9EAD3}
\begin{tabular}[c]{@{}c@{}}0.11\%\\ $\pm$0.01\end{tabular}
\\ 
 \hline
                                                                     &
\textbf{\begin{tabular}[c]{@{}r@{}}task duration (s)\end{tabular}}                                           & \textbf{0.5}                                                                                   & \textbf{1}                                                                                        & \textbf{3.5}  & \textbf{5}                                                                                                              \\ \hline
\end{tabular}
\end{table}

\subsection{Network Usage Overhead} 

As presented in Figure~\ref{fig:network_overhead}, \textbf{ProvLight transmits about 2x less data than ProvLake and DfAnalyzer.} ProvLight network usage is around 3.7~KB/sec during data capture. The application layer protocol used in ProvLight (\emph{e.g.,} MQTT-SN), which compresses captured data before transmitting it, especially for tasks with many attributes per task (\emph{e.g.,} 100 in this case), explains such difference (2x less data) when compared to the other capture approaches.

\subsection{Power Consumption Overhead} 

\begin{figure*}[t]

\begin{subfigure}{.25\textwidth}
  \centering
  \includegraphics[width=\linewidth]{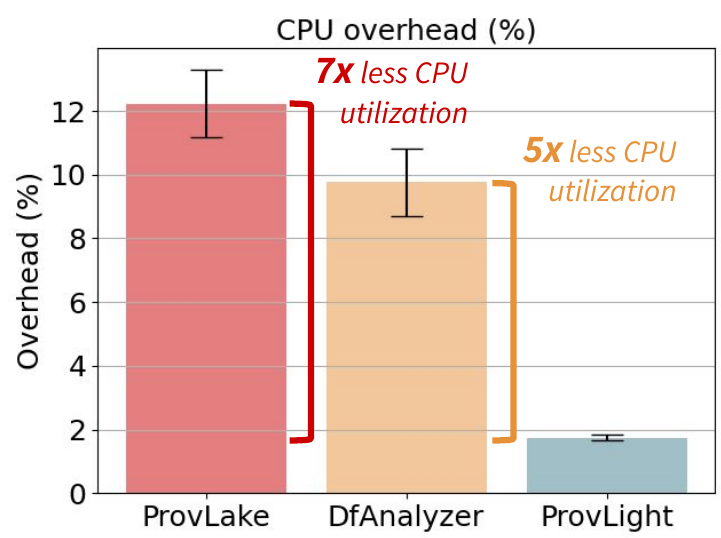}
  \caption{CPU overhead.}
  \label{fig:cpu_overhead}
\end{subfigure}%
\begin{subfigure}{.25\textwidth}
  \centering
  \includegraphics[width=\linewidth]{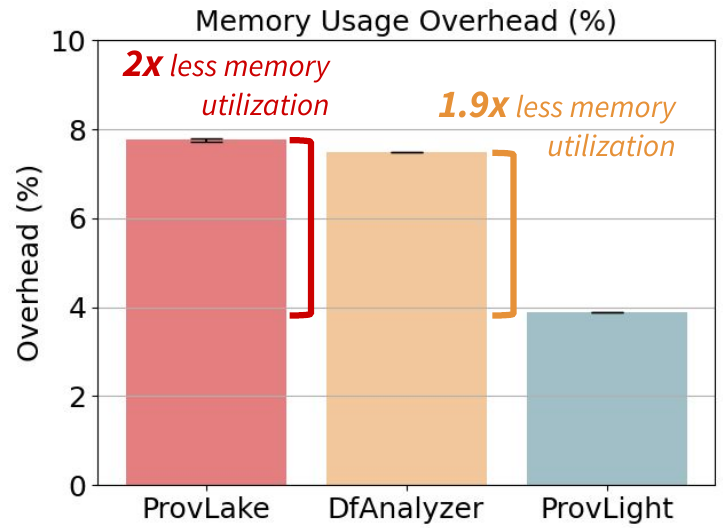}
  \caption{Memory overhead.}
  \label{fig:memory_overhead}
\end{subfigure}%
\begin{subfigure}{.25\textwidth}
  \centering
  \includegraphics[width=\linewidth]{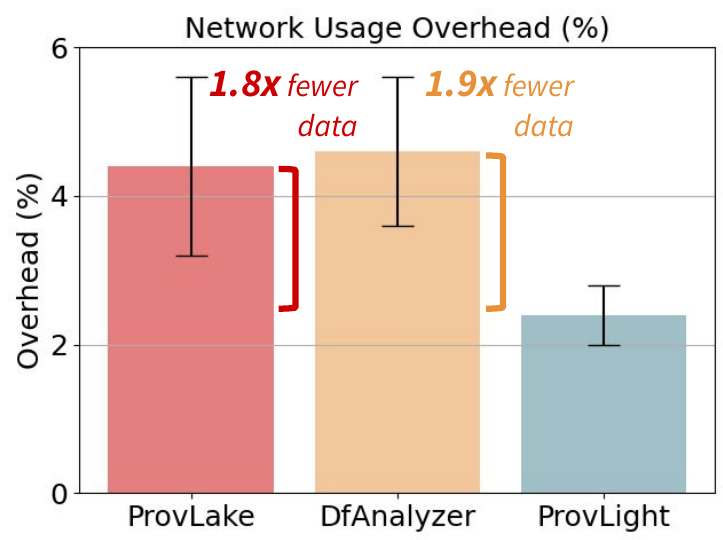}
  \caption{Network overhead.}
  \label{fig:network_overhead}
\end{subfigure}%
\begin{subfigure}{.25\textwidth}
  \centering
  \includegraphics[width=\linewidth]{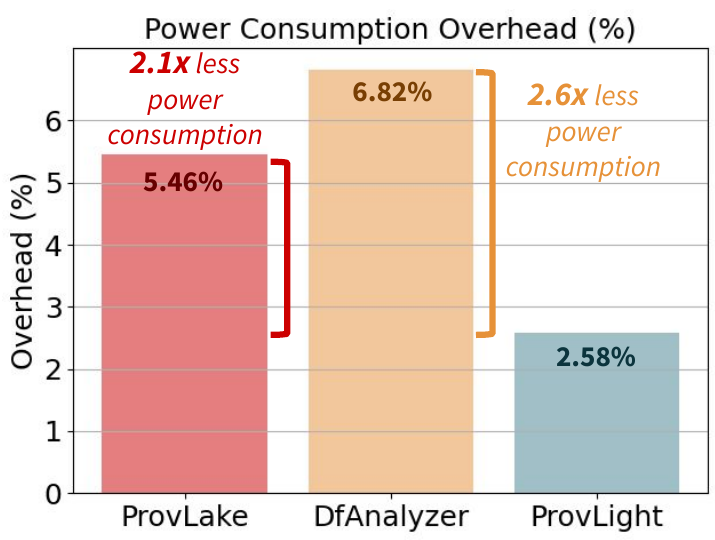}
  \caption{Power overhead.}
  \label{fig:power_overhead}
\end{subfigure}%
\caption{Provenance data capture overhead with respect to: CPU, memory, network usage, and power consumption.}
\label{fig:overhead_analysis}
\end{figure*}

Finally, results in Figure~\ref{fig:power_overhead} (error bar omitted because we use the maximum power consumption for capturing provenance data) show that ProvLight power consumption overhead is 2.1x and 2.6x less than ProvLake and Dfanalyzer. We highlight that \textbf{ProvLight overhead is 2.58\% (considered low, $<$3\%), against 5.46\% (ProvLake) and 6.8\% (DfAnalyzer).} The power consumption (in watts) for capturing and transmitting the data is on average 1.43W, 1.47W, and 1.49W for ProvLight, ProvLake, and DfAnalyzer, respectively.


\subsection{Performance in Cloud Servers}

We compare the capture time overhead of ProvLight against ProvLake and DfAnalyzer in Cloud servers (\emph{i.e.,} data capture on a server~\cite{grosg5k} available in Grid'5000). Experiment results in Table~\ref{tbl:overhead_comparison_cloud} show that the three approaches present low capture overhead ($<$3\%) for all task durations. Similarly to IoT/Edge devices, ProvLight also outperforms ProvLake and DfAnalyzer in Cloud servers. ProvLight is 7x and 5x faster than ProvLake and DfAnalyzer, respectively. ProvLight capture time overhead is very low ($<$0.25\%) for all task durations.

\section{Discussion}
\label{sec:discussion}

The integration of ProvLight as a key system within the E2Clab framework exhibits a series of features that make E2Clab a promising platform for future performance optimization of applications on the Edge-to-Cloud Continuum through efficient provenance capture and reproducible experiments.


\subsection{ProvLight Design Choices Impact on Performance}

As presented in Table~\ref{tbl:provlight-details}, the combination of ProvLight design choices on the server and client sides contributed to the low capture overhead. The ProvLight client library keeps the connection to the remote server open while capturing data (\emph{i.e.,} when capturing data from different tasks, the connection is reused). Additionally, the library is based on the publish/subscribe asynchronous communication model and it uses MQTT-SN (application layer protocol) over UDP (transport layer protocol) instead of HTTP over TCP. Despite TCP being more reliable (\emph{e.g.,} uses acknowledgment messages for data delivery), the ProvLight client sends data using QoS level 2, which guarantees that each message is received exactly once by the recipient. Such design choices help to reduce connection overheads while data transmission handshakes/acknowledgments require less bandwidth.

Another important feature is that ProvLight compresses data (using binary format) before transmitting. Through preliminary experiments, we analyzed the performance trade-offs of compressing the data on the IoT/Edge devices to make sure it is worth adding that feature. The time required to compress data (\emph{e.g.,} tasks with 100 attributes) on the edge device is negligible, around 0.001s on average. 

Our analysis considered low-bandwidth scenarios and also the data grouping strategy, resulting in fewer and larger messages to reduce the number of transmissions. We also observe that the overhead of decompressing and translating such data on the Cloud server is negligible, around 0.005s.

Data communication is key to performance efficiency in IoT/Edge workloads, especially for low bandwidth networks. ProvLight design choices such as simplified capture library for provenance data exchange (see Table~\ref{tbl:provlight-w3c}), asynchronous MQTT-SN over UDP, data grouping, and data compression, explain the positive effects on performance and costs (\emph{e.g.,} lower overheads in terms of data capture time, and CPU, memory, network usage, and energy consumption).

In summary, the lightweight asynchronous protocol (MQTT-SN over UDP) has a major impact on the capture time overhead, energy consumption, and CPU and network usage. Our simplified data model has a major impact on memory consumption, and it helps to reduce even more the capture time overhead and CPU usage by 1.7\% and 1.4\%, respectively.







\subsection{Impact of ProvLight on Real-life Use-Cases}



To illustrate how real-life use cases could benefit from ProvLight and its integration in the E2Clab framework, we consider the training of Neural Networks presented in~\cite{pina2021provenance} and~\cite{silva2021capturing}. In these articles, the authors use the storage and query components of DfAnalyzer to store captured data during model training executed on Cloud/HPC infrastructure and then query the data. They demonstrate how provenance data may be used to answer queries like the ones we presented in Section~\ref{sec:introduction}.


Since modern AI workflows are being \textbf{executed on hybrid infrastructures}, we may instantiate this use-case (Neural Network training on the Cloud/HPC) to the context of hybrid Edge-to-Cloud Federated Learning Neural Network training. In this hybrid context, the model is now trained on various resource-limited Edge devices. Thanks to the efficient capture approach of ProvLight, users may still track the model training by capturing provenance data. Without ProvLight, capturing provenance data of this use-case on the IoT/Edge is \textbf{prohibitive due to the high overheads} imposed by the existing approaches, as presented in Section~\ref{sec:soalimitations}.

Finally, thanks to the E2Clab framework, users may easily set up the Federated Learning Neural Network training and deploy it on distributed Edge devices (to train the model) and on the Cloud server (to update the global model). Furthermore, the E2Clab \emph{Provenance Manager} allows users to store data captured with ProvLight and query them using DfAnalyzer. Therefore, through the E2Clab \emph{Provenance Manager}, users may answer the same queries mentioned earlier. We highlight that this Neural Network use case is just one example from various that could benefit from this work.

\subsection{Integration with Existing Systems} 

ProvLight is designed to be easily integrated with existing provenance systems (\emph{e.g.,} ProvLake, DfAnalyzer, PROV-IO, among others) and workflow management systems and deployment frameworks (\emph{e.g.,} Pegasus, E2Clab, among others). Such integration would enable these systems to capture provenance data (with low capture overheads) in IoT/Edge devices.

As presented in Subsection~\ref{subsec:provlight_arch}, this is possible thanks to the ProvLight \emph{provenance data translator}. It translates from the ProvLight data format to the data format of the target system. This requires users to extend the ProvLight translator. In this work, we demonstrate in Section~\ref{sec:provlight_e2clab_extension}: \emph{(i)} the integration of ProvLight with the open-source DfAnalyzer provenance system as a solution for provenance capture on the IoT/Edge; and then \emph{(ii)} we integrate this capture solution within the E2Clab framework (the \emph{Provenance Manager}) to enable provenance capture of Edge-to-Cloud workflows.

\subsection{Reproducibility and Artifact Availability}

The experimental evaluations presented in this work follow a rigorous methodology~\cite{rosendo:hal-02916032} to support reproducible Edge-to-Cloud experiments on large-scale testbeds (\emph{e.g.,} Grid'5000 and FIT IoT LAB used in our experiments). This guided us to systematically define the experimental environment (\emph{e.g.,} computing resources, services/systems, network, and application execution) through well-structured configuration files. The experiment artifacts and results  are available at~\cite{exp-artifacts}.

\section{Related Work}
\label{sec:related-work}


Tanaka et al.~\cite{tanaka2022automating} extend the Pegasus~\cite{deelman2015pegasus} Workflow Management System to support Edge-to-Cloud workflows. The paper explores performance trade-offs in managing and executing Edge-to-Cloud workloads. Pegasus provides provenance data collection capabilities to capture performance metrics during workflow execution. We highlight that Pegasus (and other systems like Kepler~\cite{ludascher2006scientific}, Taverna~\cite{oinn2006taverna}, \emph{etc.}) explores the \textbf{predefined} provenance capture approach. Pegasus automatically logs provenance data about the local execution of the application codes, such as launching them and capturing the exit status and runtime information~\cite{deelman2006managing}. While ProvLight and the systems we compared with (see Table~\ref{tbl:limitations}) explore the \textbf{user-defined} capture approach, \emph{i.e.,} the user defines what to capture by workflow code instrumentation. Furthermore, unlike ProvLight, Pegasus does not explore IoT/Edge protocols to transfer the captured data nor provides features like simplified data models and compressing and grouping messages. This may result in higher overheads compared to ProvLight, as presented in Section~\ref{sec:soalimitations}. Finally, the authors do not analyze the energy consumption of their capture approach.

A provenance collection framework for the IoT/Edge devices is proposed in~\cite{nwafor2017towards}. The proposed framework follows PROV-DM recommendations and provides provenance collection capabilities for IoT/Edge devices. Unlike our work, the authors do not validate their approach on real-life Edge devices. Also, no performance evaluations are presented to understand capture overheads.

Genoma, a distributed provenance-as-a-service system across IoT/Edge devices and Cloud servers, is proposed in~\cite{narendra2019genoma}. Genoma transmits provenance data to the Cloud using the MQTT protocol. Data is transmitted based on storage availability on the Edge device and the frequency of data communication. The authors do not evaluate the performance of Genoma. Capture overheads regarding capture time, network usage, energy consumption, and CPU and memory usage are left for future work. In contrast, ProvLight is evaluated on all the metrics mentioned above.


\section{Conclusion}
\label{sec:conclusions}

The integration of ProvLight within E2Clab makes the latter, to the best of our knowledge, the first framework to support the end-to-end provenance capture of Edge-to-Cloud workflows with low overheads across the Computing Continuum. ProvLight and E2Clab are available as open-source tools. 
In future work, we will enable the provenance capture of workflows developed in C/C++ (not only in Python) and secure the data transmission from the Edge devices to the provenance system.

\section*{Acknowledgments}

This work was funded by Inria through the HPC-BigData Inria Challenge (IPL), by the French ANR OverFlow project (ANR-15- CE25-0003), and the HPDaSc associate team with Brazil. Marta Mattoso and Débora Pina are funded by CNPq and FAPERJ.
Renan is at the Oak Ridge Leadership Computing Facility at the Oak Ridge National Laboratory, which is supported by the Office of Science of the U.S. Department of Energy under Contract No. DE-AC05-00OR22725.
Experiments presented in this paper were carried out using the Grid'5000 and FIT IoT LAB testbeds, supported by a scientific interest group hosted by several Universities and organizations.

\balance
\bibliographystyle{IEEEtran}
\bibliography{references}

\end{document}